\documentclass[pra,preprint,showpacs]{revtex4}
\usepackage{amsmath}
\usepackage{graphicx}
\usepackage{epsfig}
\usepackage{amsfonts}

\def\beq{\begin{eqnarray}}
\def\eeq{\end{eqnarray}}
\def\bseq{\begin{subequations}\begin{eqnarray}}
\def\eseq{\end{eqnarray}\end{subequations}}

\begin{document}

\title{Dark states of a moving mirror in the single-photon strong-coupling regime}
\author{Gao-Feng Xu}
\affiliation{Department of Physics and Institute of Theoretical
Physics, The Chinese University of Hong Kong, Shatin, Hong Kong
Special Administrative Region, People's Republic of China}
\author{C. K. Law}
\affiliation{Department of Physics and Institute of Theoretical
Physics, The Chinese University of Hong Kong, Shatin, Hong Kong
Special Administrative Region, People's Republic of China}

\begin{abstract}
We investigate an optomechanical system in which a cavity with a
moving mirror is driven by two external fields. When the field
frequencies match resonance conditions, we show that there exists
a class of dark states of the moving mirror in the single-photon
strong-coupling regime. These dark states, which cause the cavity
to be decoupled from the external fields, is a manifestation of
quantum coherence associated with the mirror's mechanical degrees
of freedom. We discuss the properties of the dark states and
indicate how they can be generated by optical pumping due to the
decay of cavity field.

\end{abstract}

\pacs{42.50.-p, 42.50.Gy, 42.50.Wk}

\maketitle

\section{Introduction}

Quantum effects of a moving mirror interacting with
electromagnetic fields in a cavity via radiation pressure has been
a subject of considerable research interest recently
\cite{KippenbergRev,MarquardtRev,Review-Karrai,Review-Aspelmeyer,Meystre}.
The motivation of this research is not only because the
optomechanical coupling can provide novel applications in  quantum
information processing, such as the storage of optical information
as a mechanical excitation \cite{wang} and optomechanical
transducers for long-distance quantum communication
\cite{zoller_transducer}, but also because the system could be a
platform to explore fundamental quantum phenomena at macroscopic
scales. These include, for example, quantum entanglement
\cite{Tombesi,Vedral2006,Vitali2007,Paternostro2007,Hartmann2008,Meystre2},
Schr\"odinger cat states
\cite{Bose1997,Bouwmeester2003,Kleckner,cirac2}, and the
modification of uncertainty relations due to quantum gravity
\cite{kim}.

In this paper we show how a harmonically bounded mirror can evolve
into a dark state when the cavity is driven by two laser fields at
certain resonance frequencies. Dark states are well known in a
$\Lambda-$type  three-level atom, in which a coherent
superposition of two atomic ground levels can suppress light
absorption completely. Our dark states reported in this paper
share a similar mechanism, i.e., by quantum interference an
optomechanical cavity cannot absorb photons from the external
lasers.

We note that interference effects of an optomechanical cavity
driven by two light fields have been discussed in literature
\cite{Agarwal1,Agarwal2,Sidebands_OIT,OIT_science,EIT_nature,OIT_membrane}.
By linearizing the system equations, an optomechanically induced
transparency (OIT) effect, which is an analogy of
electromagnetically induced transparency, has been studied
theoretically \cite{Agarwal1,Agarwal2,Sidebands_OIT}. Recently OIT
has been observed in experiments
\cite{OIT_science,EIT_nature,OIT_membrane}. Different from
previous theoretical work, here we focus on the single-photon
strong-coupling regime in which the radiation pressure from a
single photon can displace the mirror with a displacement
comparable to the zero-point fluctuations. In such a regime, the
usual linearized photon-phonon theory becomes inadequate because
of the significant quantum fluctuations of the fields and the
mirror. To study the physics in the strong coupling regime, one
needs to solve the full quantum dynamics beyond the linear
approximation, and some authors have reported interesting
features, such as photon blockade effect \cite{Rabl_blockade},
multiple mechanical sidebands~\cite{Girvin_strong}, single-photon
scattering~\cite{Liao2012} and cooling \cite{Girvin_cooling}.

It should be noted that in previous studies of OIT, the
transparency refers to the non-absorption of a probe field, while
the cavity contains a large number of photons due to the presence
of a control field
\cite{Agarwal1,Agarwal2,Sidebands_OIT,OIT_science,EIT_nature,OIT_membrane}.
Here we show that by exploiting the single-photon strong coupling,
the driven cavity can have zero photon when the mirror is in the
dark state. In this paper we discuss how such dark states of the
mirror exist under a certain rotating wave approximation, and we
indicate how they can be prepared by an optical pumping effect.

\section{The model}

We consider an optomechanical cavity formed by a harmonically
bounded movable end-mirror and a fixed end-mirror [Fig.1(a)], in
which the cavity field and the movable end-mirror are coupled with
each other via radiation pressure. The optomechanical cavity is
driven by two lasers with frequencies $\omega_{1}$ and
$\omega_{2}$. The Hamiltonian of the system is given by
\begin{equation}
 H = \omega_c a^\dag  a + \omega _M b^\dag  b
- ga^\dag  a\left( {b^\dag   + b} \right)+\left[ {\left( {\Omega _1
e^{ - i\omega _{1} t}  + \Omega _2 e^{ - i\omega _{2} t} }
\right)a^\dag   + h.c.} \right]
\end{equation}
where $a$ $(b)$ and $\omega_c$  ($\omega_M$) are respectively the
annihilation operator and resonant frequency of the cavity field
(mechanical) modes. The $\Omega_1$ and $\Omega_2$ are proportional
to the amplitudes of the external fields. The parameter $g$ is the
single-photon coupling strength between the intracavity photon and
the mirror caused by radiation pressure. In the frame rotating at
the frequency $\omega_c$, the transformed Hamiltonian is:
\begin{equation}
 H_r = \omega _M b^\dag  b - ga^\dag  a\left(
{b^\dag   + b} \right) + \left[ {\left( {\Omega _1 e^{ - i\Delta _1
t}  + \Omega _2 e^{ - i\Delta _2 t} } \right)a^\dag + h.c.} \right]
\end{equation} where the detunings $\Delta_1=\omega _{1}-\omega_c$ and $\Delta_2=\omega
_{2}-\omega_c$  are defined.

The first two terms of $H_r$ corresponds to the Hamiltonian $H_0$
without driving, and it can be diagonalized by introducing a
displaced oscillator basis as: \begin{equation} H_0= \omega _M
b^\dag b - ga^\dag a\left( {b^\dag + b} \right) =
\sum\limits_{n,p} {\varepsilon _{n,p} \left| {\psi _{n,p} }
\right\rangle \left\langle {\psi _{n,p} } \right|}.
\end{equation}
Here the eigenvectors $\left| {\psi _{n,p} } \right\rangle  =
\left| n \right\rangle_c \otimes D\left( {ng/\omega _M }
\right)\left| p \right\rangle_M= \left| n \right\rangle_c \otimes
\left| {\tilde p (n)} \right\rangle _M $, with $n (p)$ being the
cavity photon (phonon) number. The $\left| {\tilde p (n)}
\right\rangle _M$ denotes the $n$-photon displaced Fock state of
the mirror via the displacement operator $D\left( {ng/\omega _M }
\right) = \exp [\frac{ng}{{\omega _m }}\left( {b^\dag - b}
\right)] $. The energy eigenvalues of $H_0$ are $\varepsilon
_{n,p} = p\omega _M  - \frac{{n^2 g^2  }}{{\omega _M }}$, which
depend nonlinearly on photon number $n$, and linearly on phonon
number $p$.

By using the eigenbasis of $H_0$, $H_r$ becomes
\begin{equation}
H_r = \sum\limits_{n,p} {\varepsilon _{n,p} \left| {\psi _{n,p} }
\right\rangle \left\langle {\psi _{n,p} } \right|} +
\sum\limits_{n,p,p'} {\left[ {A_{p,p'}^{\left( n \right)} \left(
{\Omega _1 e^{ - i\Delta _1 t} + \Omega _2 e^{ - i\Delta _2 t} }
\right)\left| {\psi _{n,p'} } \right\rangle \left\langle {\psi
_{n-1,p} } \right| + h.c.} \right]}
\end{equation}
where we have expressed the annihilation operator $a$ in the
eigenbasis as
 \beq a = \sum\limits_{n,p} {\sum\limits_{n',p'}
{\left| {\psi _{n,p} } \right\rangle \left\langle {\psi _{n,p} }
\right|a\left| {\psi _{n',p'} } \right\rangle \left\langle {\psi
_{n',p'} } \right|} }  = \sum\limits_{n,p,p'} {A^{(n)} _{p,p'}
\left| {\psi _{n - 1,p} } \right\rangle \left\langle {\psi _{n,p'} }
\right|} \eeq
with the coefficients $ A^{(n)} _{p,p'}  = \sqrt n
\left\langle {p} \right|D^\dag \left[ {\left( {n - 1}
\right)g/\omega _m } \right]D\left( {ng/\omega _M } \right)\left| p'
\right\rangle =\sqrt n \left\langle {p} \right|D\left( {g/\omega _M
} \right)\left| p' \right\rangle $. Specifically, the coefficients
are given by \beq A^{(n)} _{p,p'}  = \left\{
\begin{array}{l}
\sqrt{n} \sqrt {\frac{{p!}}{{p'!}}} e^{ - \frac{{\xi ^2 }}{2}}
\left( { - \xi } \right)^{p' - p} L_p^{p' - p} \left( {\xi ^2 } \right),p \le p' \\
\sqrt{n} \sqrt {\frac{{p'!}}{{p!}}} e^{ - \frac{{\xi ^2 }}{2}}
 {  (\xi) } ^{p - p'} L_{p'}^{p - p'} \left( {\xi ^2 } \right),p > p' \\
 \end{array} \right.
\eeq where $\xi =g/ \omega _M$ and $L_r^s(x)$ are associated
Laguerre polynomials.

\begin{figure}[ptb]
\center
\includegraphics[width=4.0 in]{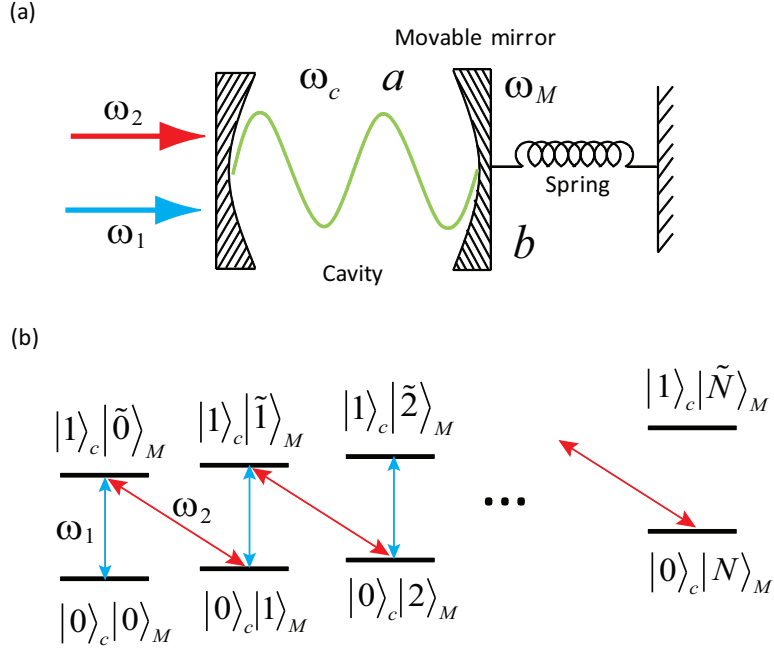}
\caption{(Color online) (a) Schematic diagram of an optomechanical
system consisting of a fixed end mirror and a movable end mirror
with two driving fields. (b) The coupling scheme between energy
levels of the optomechanical system (only zero- and one-photon
states are shown). Here each laser is used to establish a set of
resonant transitions, and $\left| {p} \right \rangle _M =\left|
{\tilde p (0)} \right\rangle _M$ and $\left| {\tilde p}
\right\rangle _M= \left| {\tilde p (1)} \right\rangle _M$ for
simplicity. By choosing $g=g_N$, there is no transition between
$|0\rangle_c |N \rangle_M$ and $|1\rangle_c | \tilde N
\rangle_M$.}\label{setup}
\end{figure}

\section{Effective resonant Hamiltonian with finite dimensions }

In this section, we show how evolution of the system state can be
confined to a finite dimensional subspace involving only the zero
and one cavity photon number and $N$ displaced phonon number
states, assuming the initial state is the ground state. This is
achieved by exploiting resonances and specific values of
optomechanical coupling strength $g$. First of all, we choose the
frequencies $\omega_1$ and $\omega_2$ of the driving fields such
that the detunings $\Delta_1$ and $\Delta_2$ satisfy the resonance
conditions:
\begin{eqnarray}
&&  \Delta _1  = \varepsilon _{1,p} - \varepsilon _{0,p} =- g^2
/\omega _M \\
&& \Delta _2  = \varepsilon _{1,p}  - \varepsilon _{0,p + 1}  =-
\omega _M  - g^2 /\omega _M
\end{eqnarray}
In this way the driving field $\Omega_1$ can resonantly couple the
states $\left| {\psi _{0 ,p} } \right\rangle$ and $\left| {\psi
_{1 ,p} } \right\rangle$ as illustrated by vertical arrows in Fig.
1b. Similarly, the driving field $\Omega_2$ can resonantly couple
the states  $\left| {\psi _{0 ,p+1} } \right\rangle$ and $\left|
{\psi _{1 ,p} } \right\rangle$.

It is important to note that $\varepsilon _{n,p}$ depends
nonlinearly on the photon number $n$, therefore the above
resonance relations do not hold for states with photon numbers
larger than $1$. In fact, if the optomechanical coupling strength
$g$ is sufficiently strong, the conditions (7) and (8) correspond
to far off resonance for transitions from 1-photon states to
2-photon states, and this leads to the photon blockade effect
\cite{Rabl_blockade}. Hence if the driving fields are sufficiently
weak, it is justified to ignore the off-resonant transitions from
1-photon states to 2-photon states. Specifically, we require (for
$i=1,2$),
\begin{equation}
\Omega _i \ll \left|2g^2 /\omega _M
- { K\omega _M } \right|
\end{equation}
where $K$ is the nearest integer to $2 (g /\omega _M)^2$. For
example, if $g < \omega_M /2$ then $K=0$. The inequality (9) means
that $\Omega_i$ should be much less than the smallest detuning
between 2-photon manifold to 1-photon manifold.

With the conditions (7-9), if the initial cavity photon number is
zero, the system can be effectively confined to the subspace of
zero and one photon. Furthermore, we will keep only the
co-rotating terms in (4) as a rotating wave approximation. Under
such an approximation, $\Omega_i$ can drive the corresponding
resonant transitions only, and the Hamiltonian (4) in the
interaction picture [i.e., the first term in (4) is removed by a
rotating frame] becomes,
\begin{eqnarray}
 H_{r}'  =
 \sum\limits_{p = 0}^{N - 1} {\left[ {\left( {A_{p,p}^{\left( 1 \right)}
  \Omega _1  \left| {\psi _{1,p} } \right\rangle
  \left\langle {\psi _{0,p} } \right| + A_{p + 1,p}^{\left( 1 \right)}
  \Omega _2  \left| {\psi _{1,p} } \right\rangle
  \left\langle {\psi _{0,p + 1} } \right|} \right) + h.c.}
  \right]}.
 \end{eqnarray}
Here the upper limit of phonon number $N$ can be infinite in
general. However, a {\em finite} value of $N$ is possible if the
transition between $\left| {\psi _{0 ,N} } \right\rangle$ and
$\left| {\psi _{1 ,N} } \right\rangle$ can be completely
suppressed (Fig. 1b). This can be achieved by a specific value of
$g=g_N$ such that
\begin{equation}
A_{N,N}^{\left( 1 \right)}= \exp ( - \frac{{g_N ^2 }}{2
\omega_M^2}) L_N^0 \left( {g_N^2/\omega_M^2 } \right) = 0.
\end{equation}
In other words, one can design the Hamiltonian $H_r'$ with a
prescribed $N$ by using the optomechanical coupling strength
$g_N$. In Fig. 2, we illustrate the roots $g_N$ as a function of
$N$.  We see that the value of $g_N$ decreases when $N$ becomes
larger and larger. For example, $g_N \approx 0.12 \omega_M$ for
$N=100$.

\begin{figure}[ptb]
\center
\includegraphics[width=3.4 in]{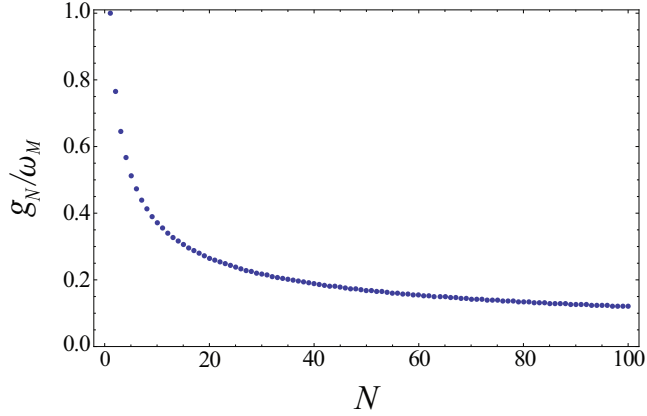}
\caption{(Color online) Solution of $g_N$ satisfying Eq. (11) as a
function of $N$. For each $N$, there are multiple roots, and only
the smallest positive root is shown.}\label{setup}
\end{figure}

\section{Dark states}


The Hamiltonian $H_r'$ has an eigenvector $|D \rangle$ with a zero
eigenvalue:
\begin{equation} \left| {D
} \right\rangle  = C \sum\limits_{p = 0}^N \beta_p | p \rangle _M
\otimes \left| 0 \right\rangle _c
\end{equation}
where $\beta_0=1$, and for $p>0$,
\begin{equation}
\beta_p =  ( { - 1} )^p \left( {\frac{{\Omega _1 }}{{\Omega _2 }}}
\right)^p {\prod\limits_{i = 0}^{p - 1} {\frac{{A_{i,i}^{\left( 1
\right)} }}{{A_{i+1,i}^{( 1 )} }}} }
\end{equation}
and $C$ is a normalization constant. Such an eigenvector is a
coherent superposition of phonon states and the cavity contains no
photon. It is a dark state induced by quantum coherence of the
mirror, and the interference forbids any excitation of cavity
field even though the cavity is constantly driven by the two
external fields.

\begin{figure}[ptb]
\center
\includegraphics[width=3.2 in]{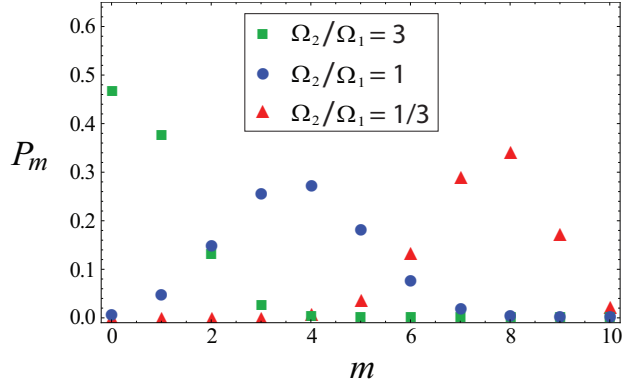}
\caption{(Color online) Phonon number distributions of dark states
for different ratios of driving strengths at the coupling strength
$g=g_{10}=0.37\omega_M$. }\label{setup}
\end{figure}

\begin{figure}[ptb]
\center
\includegraphics[width=3.2 in]{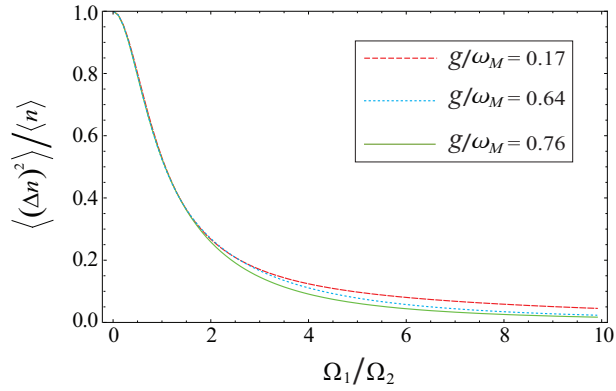}
\caption{(Color online)  $\langle(\Delta n)^2\rangle/\langle
n\rangle$ as a function of $\Omega_1/\Omega_2$ for $g/ \omega_M=
0.17, \ 0.64, \ 0.76$ corresponding to $N=20,\ 3,\ 2$
respectively. }\label{setup}
\end{figure}

The phonon number distribution of the dark states is complicated
by the Laguerre functions in Eq. (13). While the details form of
$|\beta_p|^2$ requires a numerical evaluation of Eq. (13), we find
that $|\beta_p|^2$ is mainly controlled by the ratio of the
strengths of driving fields. Such a feature is illustrated in Fig.
3 for various $\Omega_2/\Omega_1$. For example, when
$\Omega_2/\Omega_1= 3$, the probability decreases quickly with the
increase of phonon number $m$. In the case $\Omega_2/\Omega_1= 1$,
there is a peak appears in the probability distribution. If the
ratio is further decreased to $\Omega_2/\Omega_1= 1/3$ , the peak
is shifted towards higher phonon numbers.

It is worth noting that phonon number distributions of dark states
exhibit a sub-Poissonian statistics. This is illustrated in Fig. 4
in which the ratio $\langle(\Delta n)^2\rangle/\langle n\rangle$
as a function of $\Omega_1/\Omega_2$ is plotted. We see that
$\langle(\Delta n)^2\rangle/\langle n\rangle$ is always less than
1 (i.e., sub-Poisson distribution) except at the small region
$\Omega_1/ \Omega_2 $ near zero. In particular, the
$\langle(\Delta n)^2\rangle/\langle n\rangle$ decreases with
$\Omega_1/ \Omega_2$. In Fig. 4, we also find that the curves are
quite insensitive to the value of $g_N$ used.

\section{Preparation of dark states by cavity-field damping}

In this section we discuss how the system can be optically pumped
into the dark state by cavity-field damping. The evolution of the
dissipative system is governed by the master equation:
\begin{equation}
\frac{{d\rho }}{{dt}} =  - i\left[ {H_r,\rho } \right] -
\frac{{\gamma _c }}{2}\left( {a^\dag a\rho  - 2a\rho a^\dag   +
\rho a^\dag  a} \right) \end{equation} where $\rho$ is the density
matrix of the photon-mirror system, and $\gamma_c$ is the
cavity-field decay rate. Here we have assumed that the mechanical
motion of the mirror has the damping rate $\gamma_M$ that is  much
smaller than $\gamma_c$. In addition, we will focus on a finite
time interval $ 1/ \gamma_c \ll t \ll 1/\gamma_M$ during which the
optically pumping is complete while the decay of mirror's motion
remains negligible, and so we only include the decay effect of the
cavity field in the above master equation. Note that the original
Hamiltonian $H_r$ defined in Eq. (2) is used without employing the
rotating wave approximation in Eq. (10). If $H_r$ is simply
replaced by $H_r'$, then $\rho = |D \rangle \langle D|$ is already
a steady state solution of the master equation Eq. (14), because
the cavity field damping term has no effect on $|D \rangle$ (which
has zero photon).

We have solved the master equation (14) numerically with an initial
ground state of the system. Specifically, we are interested in the
fidelity $F$ defined by
\begin{equation}
F = Tr\left( {\left| {D} \right\rangle \left\langle {D}
\right|\rho \left( t \right)} \right),
\end{equation}
which measures the probability of the system in the dark state $|D
\rangle$. Some examples are given in Fig. 5 in which $F$ increases
with time and approaches a steady value close to 1 in a finite
time. For the three cases shown in Fig. 5, the fidelities can
reach $F \approx 0.99$ with $g_N/\omega_M=0.37$ ($N=10$) at the
time $T \approx 8000 \omega_M^{-1}$. To ensure that the mechanical
decoherence is negligible, we need $\gamma_M < 10^{-4} \omega_M$
for the parameters used in Fig. 5. Indeed, we have tested
numerically the performance by including mechanical damping in the
master equation, and found that $F \approx 0.99$ when $\gamma_M
=10^{-5} \omega_M$ and $F \approx 0.93$ for a larger $\gamma_M
=10^{-4} \omega_M$. We remark that the time $T$ required to
generate dark states is shorter for smaller $N$. For example, when
$N=3$, our numerical calculations indicate that $T \approx 2000
\omega_M^{-1}$.

\begin{figure}[ptb]
\center
\includegraphics[width=3.2 in]{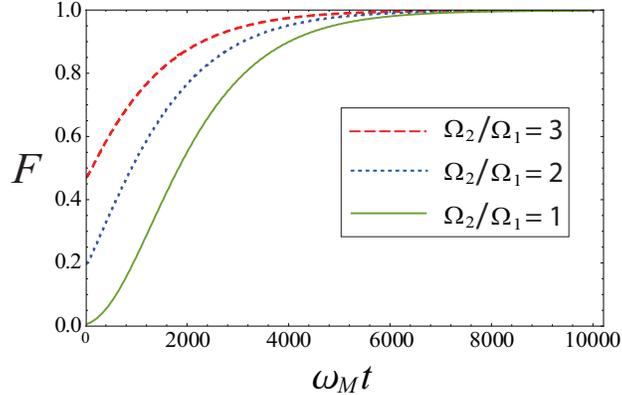}
\caption{(Color online) Time evolution of the fidelity $F$  for
various $\Omega_2/\Omega_1$ ratios. The parameters are: $N=10$,
$g/\omega_M=0.37$, $\gamma_c/\omega_M=0.05$, $\Omega _2 /\omega_M=
0.01$, $\Delta_1/\omega_M=-0.14$, $\Delta_2/\omega_M=-1.14$.}
\end{figure}

The increase of $F$ is understood as an optical pumping effect.
This is because when a photon leaks out of the cavity, the mirror
can have a non-zero probability making a transition to the dark
state. Since the dark state is decoupled from the driving fields,
it can no longer be excited, and hence the occupation of the dark
state accumulates as time increases. In our system, the loss is
mainly due to a leakage of phonon population beyond the phonon
number $N$, because the cavity field decay also causes the mirror
to make transitions to states other than the dark state. Note that
the transition amplitudes from $|1 \rangle_c |\tilde p \rangle_M$
to $|0 \rangle_c |q \rangle_M$ due to the transmission of a photon
out of the cavity is proportional to the Frank-Condon factor
$\langle p | \tilde q \rangle_M$, the loss can be reduced by
choosing a sufficiently high $N$ and $\Omega_2> \Omega_1$. This is
because dark states with $\Omega_2> \Omega_1$ concentrate on lower
phonon numbers (Fig. 3) and hence the Frank-Condon factors for
transitions to states of phonon number higher than $N$ is smaller.

\section{Conclusion Remarks}

To conclude, we have addressed a quantum interference effect in
the single-photon strong-coupling regime of optomechanics. In such
a regime we discover a class of dark states of a moving mirror
under the conditions (7-9). These dark states make the cavity
decoupled from two external driving fields, and the decoupling is
derived without employing the linearization treatment. We provide
an analytical expression of the dark states which indicate the
dependence of the ratio $\Omega_2/\Omega_1$ of the driving fields
and the optomechanical coupling strength $g$. With the help of
cavity damping,  dark states can be prepared by optical pumping.

In this paper specific optomechanical coupling strengths $g_N$
have been used in order to `trap' the mirror state in the space of
finite phonon numbers. It is important to ask how a slight
deviation of $g$ away from $g_N$ would affect the dark state
generation, because the trapping effect would become imperfect
when $g \ne g_N$. To address the issue, we have tested the
sensitivity of the fidelity to small variations of $g$ values. For
example, with $\Omega_2 > \Omega_1$ and parameters in Fig. 5, by
introducing a 3$\%$ deviation of the $g$ value we found that the
fidelities can still reach about $F \approx 0.99$. This can be
understood by Eq. (13). Strictly speaking, when $g$ is not exactly
equal to $g_N$, the upper limit $N$ in Eq. (12) should be extended
to infinity because $A^{(1)}_{NN}$ is no longer zero. However, if
$\Omega_1/ \Omega_2$ is less than one, then $\beta_p$ appearing in
Eq. (13) can quickly converge to zero, yielding a significant
population concentrated at phonon numbers much lower than $N$
(Fig. 3). In this case the leakage of phonon numbers beyond $N$
would be negligible and hence  a slight deviation of $g$ would not
affect the dark state generation appreciably. On the other hand,
we find that the system with $\Omega_2 < \Omega_1$ is quite
sensitive to $g$. This is because if $\Omega_1/ \Omega_2 > 1$,
$\beta_p$ in Eq. (13) tends to shift to higher phonon numbers
(Fig. 3), making the leakage beyond $N$ significant. Therefore it
would be more favorable to employ $\Omega_2
> \Omega_1$ for experimental realizations of our scheme.  Finally,
we remark that the realization of single-photon strong-coupling
regime is still a challenge for current experiments, but some
recent progress have been made in various optomechanical systems
in order to reach the strong coupling regime
~\cite{Gupta2007,Brennecke2008,Eichenfield2009}.

\begin{acknowledgments}
The authors would like to thank Jieqiao Liao for discussions. This
work is supported by the Research Grants Council of Hong Kong,
Special Administrative Region of China (Project No.~CUHK401810).
\end{acknowledgments}

\end{document}